\documentclass[preprint]{aastex}

\usepackage{natbib}
\newcommand{\teff}{T_{\rm{eff}}}
\newcommand{\numax}{\nu_{\rm{max}}}

\newcommand{\numod}{\nu_{\rm{mod}}}
\newcommand{\feh}{\rm{[Fe/H]}}

\newcommand{\Dnu}{\Delta\nu}
\newcommand{\chired}{\chi^2_{\nu}}
\newcommand{\be}{\begin{equation}}
\newcommand{\ee}{\end{equation}}
\newcommand{\bea}{\begin{eqnarray}}
\newcommand{\eea}{\end{eqnarray}}
\newcommand{\msun}{M$_\odot$}
\newcommand{\rsun}{R$_\odot$}

\received{}
\accepted{}

\slugcomment{}
\shorttitle{Modeling the surface term}
\shortauthors{Schmitt \& Basu}

\begin{document}
\title{Modeling the Asteroseismic Surface Term across the HR Diagram}
\author{
Joseph R. Schmitt\altaffilmark{1} and
Sarbani Basu\altaffilmark{1}
} 
\email{joseph.schmitt@yale.edu}

% Some possible altaffiltext entries
\altaffiltext{1}{Department of Astronomy, Yale University, New Haven, CT 06511 USA}

\begin{abstract}

Asteroseismology is a powerful tool that can precisely characterize the mass, radius, and other properties of field stars.  However, our inability to properly model the near-surface layers of stars creates a frequency-dependent frequency difference between the observed and the modeled frequencies, usually referred to as the ``surface term''. This surface term can add significant errors to the derived stellar properties unless removed properly.  In this paper we simulate surface terms across a significant portion of the HR diagram, exploring four different masses ($M=0.8, 1.0, 1.2$, and $1.5$~\msun) at five metallicities ($\feh=0.5, 0.0, -0.5 ,-1.0, and -1.5$) from main sequence to red giants for stars with $\teff<6500 K$ and explore how well the most common ways of fitting and removing the surface term actually perform. We find that the two-term model proposed by \citet{Ball2014} works much better than other models across a large portion of the HR diagram, including the red giants, leading us to recommend its use for future asteroseismic analyses.

\end{abstract}


\section{Introduction}
\label{sec:intro}

Asteroseismic analysis of stars is the most robust way of determining stellar properties. Asteroseismic data allow us to determine the masses and radii of stars to exquisite precision.  NASA's \textit{Kepler} mission has provided asteroseismic data on hundreds of dwarfs and tens of thousands of giants. These data have revolutionized our knowledge of the target stars. Achievements include determining the properties of exoplanet hosts, and thereby the radii of the planets themselves \citep[e.g.,][]{Howell2012, Borucki2012, Carter2012}, and catalogs of the properties of dwarfs \citep{Chaplin2014} and giants \citep{Pinsonneault2014} in the {\it Kepler} field. We can also determine fairly precise ages for stars, albeit in a model dependent way \citep[][etc.]{Metcalfe2010, Metcalfe2012, Mathur2012, Silva2015}.

Knowledge of stellar properties, such as age, can illuminate research in a number of different fields: planet formation and migration, stellar evolution, and the star formation history and chemical evolution of the Milky Way.  The sources of age uncertainties for asteroseismology are vastly different from other methods.  The best determined ages outside of asteroseismology are fitting color-magnitude diagrams to theoretical isochrones for star clusters.  However, this has the obvious disadvantage that it cannot be applied to field stars.  Fitting isochrones to single field stars is plagued with problems, such as the ``terminal age bias'', the tendency for stars to be systematically pulled towards the end of the main sequence \citep{Pont2004, Pont2005}.  Even the more sophisticated, Bayesian models \citep{Jorgensen2005, Takeda2007} that avoid the terminal age bias often come with large uncertainties.  This is further complicated by observational difficulties, such as reddening and unknown distances.   Asteroseismology, on the other hand, is unaffected by reddening or distance.  

For most stars, we only have two pieces of asteroseismic data: the large separation, $\Dnu$, and the frequency of maximum power, $\numax$. Additionally, for many of the observed stars, precise asteroseismic data, i.e., frequencies of individual modes, have also been observed. Detailed asteroseismic analyses can be done for these stars. The usual way to determine the properties of these stars is to construct models that reproduce not just the ``classical'' properties of the star (i.e., $\teff$, metallicity, and position on the color-magnitude diagram), but also the pulsation frequencies. Since the frequencies are usually extremely precise, the stellar properties can also be determined precisely.

There is, however, one major issue that plagues asteroseismic analyses, and that is the so-called ``surface term.'' Despite improvements in stellar models, we are still unable to model the near-surface layers of a star properly. The major source of uncertainty is how we model convection.  Convection is an inherently three-dimensional, dynamical phenomenon, and we approximate that to a static, one-dimensional equation either through mixing length formalism or through other approximations, such as that of \citet{Canuto1991}. The mixing length approximation also ignores the dynamical effects of convection, such as turbulent pressure support and kinetic energy flux. Most of the effect of the lack of proper modeling lies very close to the surface of the model, notably in the so-called super-adiabatic layer, where convection is inefficient. The consequence of this is a frequency-dependent frequency difference between frequencies of the model and that of the star, which was first observed in the case of the Sun. This is usually called the ``surface term'' \citep{Christensen1991}. The surface term also depends on the atmospheric physics used to model the stars. \citet{Christensen1991} showed that if one corrected for the mode-inertia, then the frequency differences due to the imperfect modeling of the near-surface layers is a function of frequency alone.

Helioseismic analysis of the Sun proceeds through the removal of the surface term using a subset of the solar frequencies. This is possible because many thousands of solar oscillation frequencies with degrees ranging from $l=0$ (the radial modes) to about $l=300$ have been precisely determined. Filtering out the surface term in the way it is done for the Sun is not possible for other stars; this is a consequence of the fact that only low-degree mode-frequencies ($l=0,1,2$ and sometimes $l=3$) are observed for these stars. This does not give us enough information to filter out the surface term and determine other properties of the star at the same time. In the solar case, we also have the advantage that the mass, radius, and age are known independently, and thus, the seismic analysis can focus on determining the details of solar structure. For the vast majority of other stars, the mass and radius of a star have to be determined from the same data set that is used to determine age and other properties of the star, and thus, filtering out the surface term without putting any constraints on what we expect becomes impossible. The usual way out of this bind is to assume that the surface term present between the frequency differences of a star and its model can be assumed to be a scaled version of the surface term present in the frequency differences between the Sun and solar models. The problem with this, however, is that no one has tested if such scaling works for all models at different stages of evolution and  with non-solar metallicities. Improper removal of the surface term can affect the inference derived from the surface term. They can affect determined ages and internal structure \citep{SilvaAguirre2013}, estimates of the edges of convection zones  \citep[e.g.,][]{Basu1994, Monteiro1994, Basu1997, Mazumdar2012} and the position of the helium ionization zones \citep{Basu2004, Mazumdar2012, Verma2014}.

In this paper, we examine whether the usual ways of correcting for the surface term can be applied to all stars. For this, we simulate different forms of the surface term between pairs of models and determine how well we can remove those differences. The rest of the paper is organized as follows. We give examples of the solar surface term in Section~\ref{sec:solar}. The usual ways of determining the surface term are described in Section~\ref{sec:surf}. In Section~\ref{sec:sim}, we describe the different ways in which we have simulated the surface term. Our results are described in Section~\ref{sec:res}.

\section{The Solar Surface Term}
\label{sec:solar}

\begin{figure*}
		\plotone{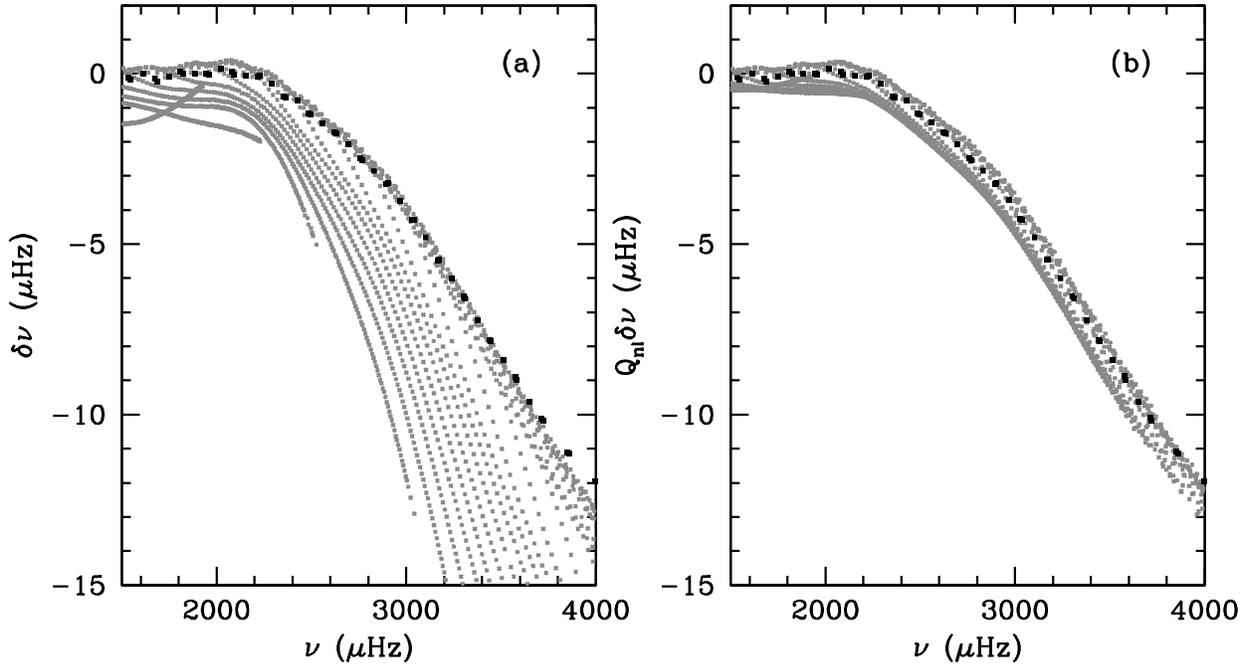}
	\caption{Frequency differences between the Sun (mode set BiSON-13 of \citeauthor{basuetal2009} \citeyear{basuetal2009}) and Model~S of \citet{Christensen1996}. Panel (a) shows the raw frequency differences, and Panel (b) the differences scaled by $Q_{nl}$. The points in black are modes of degree $l=0$--2.}
	\label{fig:modelS}
\end{figure*}

As mentioned earlier, surface terms are usually corrected by scaling the solar surface term, and thus, one cannot begin a discussion of the surface term without studying it in the Sun. In Fig.~\ref{fig:modelS}(a) we show the raw frequency differences between the Sun and the solar model known as Model S \citep{Christensen1996}, and in  Fig.~\ref{fig:modelS}(b), we show the same after the differences have been scaled by the quantity: 
\be
Q_{nl}=\frac{E_{nl}(\nu)}{E_{l=0}(\nu)},
\label{eq:qnl}
\ee
where $E_{nl}$ is the inertia of a mode of order $n$ and degree $l$, and $E_{l=0}(\nu)$ is that of an $l=0$ mode of at the same frequency as the model with order $n$ and degree $l$. This factor effectively corrects for the fact that a given perturbation changes the mode frequencies of low-inertia modes differently from that of high-inertia modes. $Q_{nl}=1$ for all $l=0$ modes by definition. Note that the scaled frequency differences are predominantly a function of frequency. The spread of the points is caused by differences in the interior, while the frequency dependence of the difference is entirely from near-surface errors. This frequency-dependent frequency difference is the surface term.

\begin{figure}
                \plotone{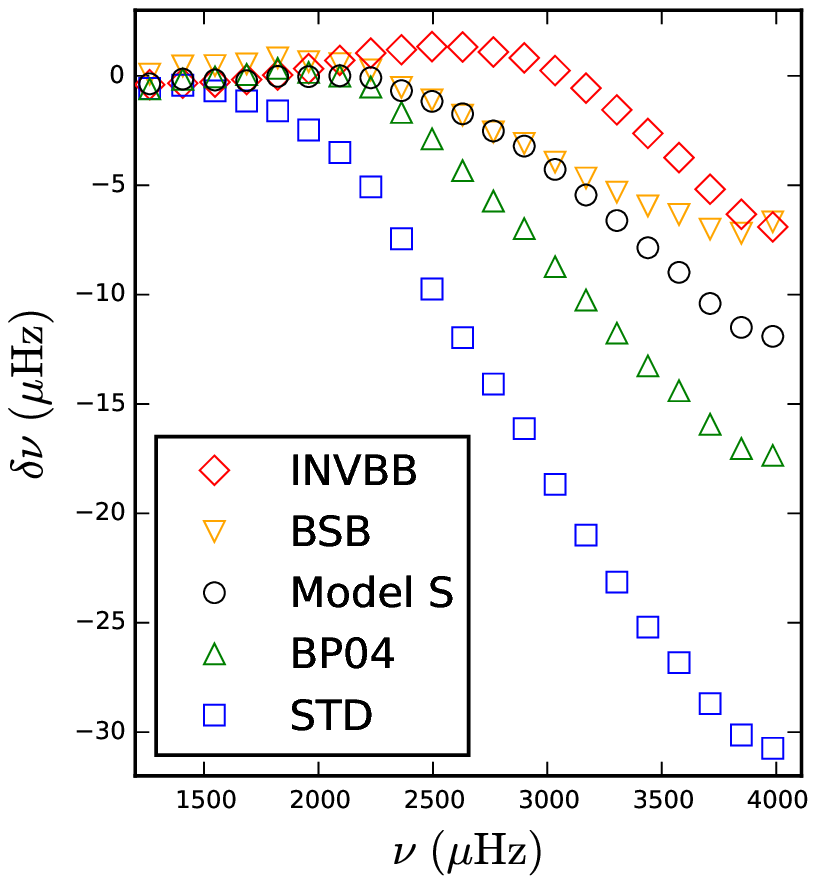}
        \caption{Frequency differences between the $l=0$ modes of the  Sun and different solar models. Model INVBB is the seismic model of \citet{antia1996}, model BSB is model BSB(GS98) of \citet{bahcalletal2006}, Model S is from \citep{Christensen1996}, BP04 is from \citet{bahcalletal2005}, and STD is from \citet{basuetal2000}.}
        \label{fig:allmodels}
\end{figure}

\begin{figure*}
                \plotone{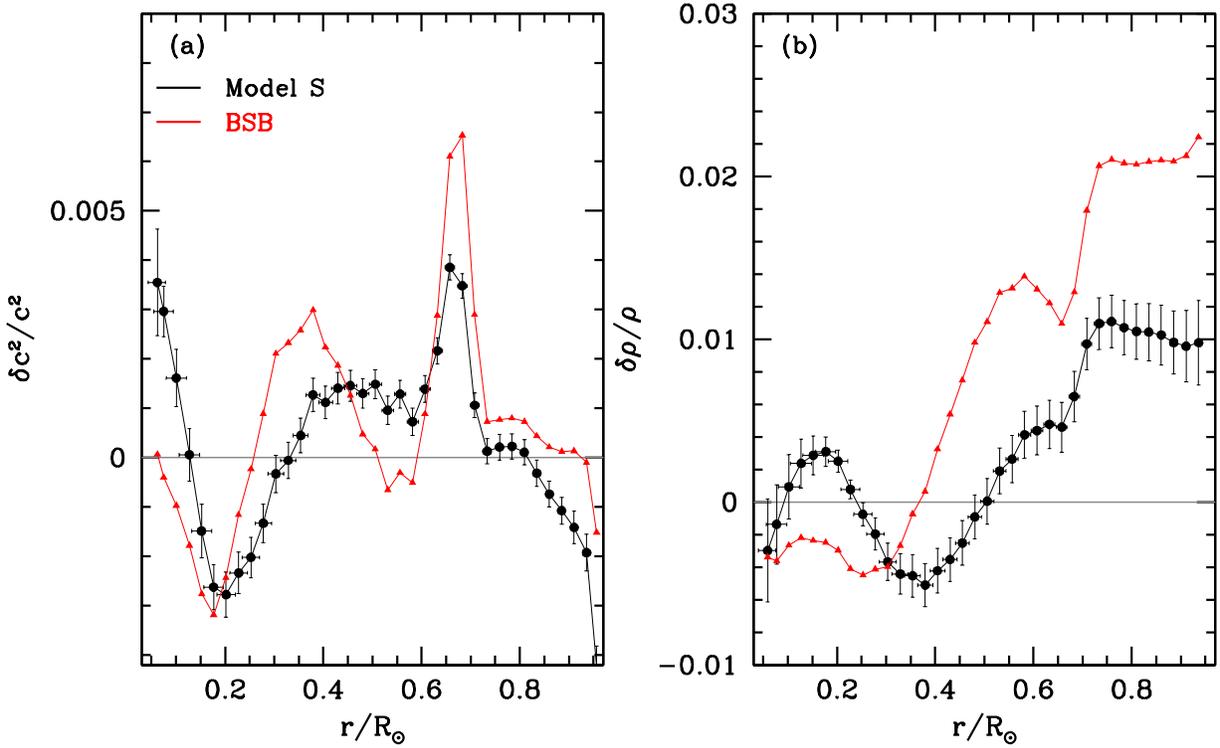}
        \caption{(a) The relative differences of the squared sound speed between the Sun and two other solar models, Model~S and model BSB. (b) The relative density differences between the Sun and the two models. The results, which  were obtained by inverting the frequency differences between the Sun and these models, are from \citet{basuetal2009}.}
        \label{fig:cdif}
\end{figure*}

For all low-degree p-modes that are observed in other stars, $Q_{nl}\simeq1$. Mixed modes, on the other hand, can have much higher values of $Q_{nl}$. Since there can be no $l=0$ mixed modes, it is often useful to define the surface term. Judging by Fig.~\ref{fig:modelS}, the surface term is quite simple: 0 at very low frequencies and then a steep change which can be modeled as a simple function, such as a power law \citep{Kjeldsen2008}. Unfortunately, however, the surface term is  not universal and depends on the physics used. This is demonstrated in Fig.~\ref{fig:allmodels}, where we show the frequency differences between the Sun and a number of published solar models. Not only do the surface terms look very different, the magnitude of the frequency difference does not tell us anything about how different the structure of the model is from the Sun (see Fig.~\ref{fig:cdif}). As is clearly seen, Model S clearly has a larger surface term than model BSB, yet in terms of structure of the interior, Model S is clearly better. Thus the surface term can mislead us about how well a model matches the star. 

Since multiple physics inputs in the models shown in Figs.~\ref{fig:allmodels}\ and \ref{fig:cdif} are  different, it is difficult to pinpoint the exact cause of the differences in the surface term. However, one can show that the surface term changes if the model atmosphere is changed, and/or near-surface opacities are changed,
 and/or the formulation of how convection is modeled is changed \citep[see][]{jcdmjt1997}. In Fig.~\ref{fig:eddiks}\ we show the frequency differences between two solar models, one constructed with the Eddington $T$-$\tau$ relation in the atmosphere and the other with the Krishna Swamy relationship \citep{KrishnaSwamy1966}. Note that the frequency differences show the typical surface term pattern. The structural differences between the two models are concentrated at the surface.  

Thus, to reiterate, the surface term, even that between the Sun and its models, is not unique. Thus, even if we  assume that the surface term of non-solar models can be corrected using the solar surface term, we first need to ensure that the models are constructed with identical physics.

\section{Usual ways of correcting for the Surface Term}
\label{sec:surf}

The most common way to correct for the surface term for stellar models is that proposed by \citet{Kjeldsen2008}. \citet{Kjeldsen2008} noted that the steep part of the surface term of Model~S could be modeled as a power law and proposed that the surface term for other stars be
corrected as:

\begin{equation}
\nu_{\rm{corr}}-\numod=a\left(\frac{\numod}{\nu_{\rm ref}}\right)^b,
\label{eq:KBCD08}
\end{equation}

\noindent where $\numod$ is the frequency of any given mode of a model, $\nu_{\rm{corr}}$ is the corrected frequency, $a$ is the correction at $\numod=\nu_{\rm ref}$, and $b$ is the slope from fitting the solar frequency differences. To avoid extra free parameters, many investigators use $\nu_{\rm ref}=\numax$ \citep[][etc.]{Mathur2012, Metcalfe2014}.  Recall that $\numax$ scales as:

\begin{equation}
{{\nu_{\rm max}}\over{\nu_{{\rm max},\odot}}}\simeq{{M/M_{\odot}}\over
 {(R/R_{\odot})^2\sqrt{(T_{\rm eff}/T_{{\rm eff},\odot})}}}
\end{equation}

\noindent \citep[e.g.][]{hans1995,bedding2003}, where $M$, $R$, and $\teff$ are the stellar mass, radius, and effective temperature.   We refer to the model in Equation~\ref{eq:KBCD08} as the KBCD08 model.   

Another way to correct for the surface term is to scale the solar surface term directly. This is applied as follows: denote the solar surface term as the  $\nu_{\odot}$-$\delta\nu_{\odot}$ relation. Both  $\nu_{\odot}$ and $\delta\nu_{\odot}$ are then scaled to the mass and radius of the stellar model under consideration using the homology scaling:

\begin{equation}
r=\frac{\Delta\nu_{\rm mod}}{\Delta\nu_\odot},
\end{equation}

\noindent where the large separation of the model $\Delta\nu_{\rm mod}$ can be obtained either from the scaling relation:

\begin{equation}
 \frac{\Delta\nu}{\Delta\nu_{\odot}}\simeq\sqrt{\frac{M/M_{\odot}}{(R/R_{\odot})^{3}}},
\end{equation}

\noindent or by calculating the large separation as an average separation from the frequencies themselves. The latter is preferred. The resulting $(r\nu_{\odot}$-$r\delta\nu_{\odot})$ relation  is then used to correct the stellar model for the surface term. The surface term of the model is then denoted as:

\begin{equation}
\nu_{\rm{corr}}-\numod=a\delta\nu_{\odot,{\rm SC}},
\label{eq:ScS}
\end{equation}

\noindent where $\delta\nu_{\odot,{\rm SC}}$ is $r\delta\nu_{\odot}$ at $r\nu_{\odot}=\numod$. We call this the Scaled Solar method, or ScS-1 (the ``1'' denoting one free parameter). As a variant, we also add a vertical offset $b$ to Eq.~\ref{eq:ScS} and call that case ScS-2. The vertical offset mimics the effect of the evolution of the
phase term in the expression for $\Delta\nu$ as a star evolves \citep[see][]{whiteetal2012}.

Since frequency changes caused by a given perturbation depends on the inertia of the mode~--- high inertia modes being perturbed less than low inertia modes --- models KBCD08, ScS-1, and ScS-2, when applied to non-radial modes, need to be scaled inversely with the $Q_{n,l}$ for that mode. This is particularly needed to ensure that mixed modes, which have very large mode inertia and often have $Q_{nl} > 1$, are handled properly.

\citet{Ball2014} have suggested a surface term correction that is completely independent of the solar surface term, which makes it very promising. It has not been widely adopted as yet, which makes testing this model more useful for all those who might wish to adopt it. \citet{Ball2014} have two  models of the surface term $\delta\nu$. They propose:

\begin{equation}
\delta\nu(\nu)=\frac{a_3}{E_{nl}}\left(\frac{\nu}{\nu_{\rm{ac}}}\right)^3,
\label{eq:bg141}
\end{equation}

\noindent where $a_3$ is a coefficient that is determined by fitting, $\nu_{\rm{ac}}$ is the acoustic-cutoff frequency (recall that $\numax\propto\nu_{\rm{ac}}$, and hence the same scaling applies), and $E_{nl}$ is the mode inertia. We call this the ``BG14-1'' model. The second model they propose is:

\begin{equation}
\delta\nu(\nu)=\frac{a_3}{E_{nl}}\left(\frac{\nu}{\nu_{\rm{ac}}}\right)^3+
\frac{a_{-1}}{E_{nl}}\left(\frac{\nu}{\nu_{\rm ac}}\right)^{-1},
\label{eq:bg142}
\end{equation}

\noindent which we call the ``BG14-2'' model. In this form there are two parameters that are determined via fitting, $a_3$ and $a_{-1}$.   The explicit dependence of expressions on the mode inertia means that that the
expressions can be applied to non-radial modes without any modification and obviates the need to scale using
$Q_{n,l}$.

\section{Simulating the Surface term}
\label{sec:sim}

\subsection{Calibrated models with different atmospheres}
\label{subsec:calib}

Given that the surface term is only known in the solar case, testing out the fitting forms requires us to simulate a known surface term between  models and the stars they represent. We are guided in this by the solar case where \citet{jcdmjt1997} showed that changing the model atmosphere, surface opacities, and/or the mixing length formalism can simulate a surface term. Since changes in the model atmosphere produce a surface term between solar models, we use this to construct surface terms for other models too.

\begin{figure}
                \plotone{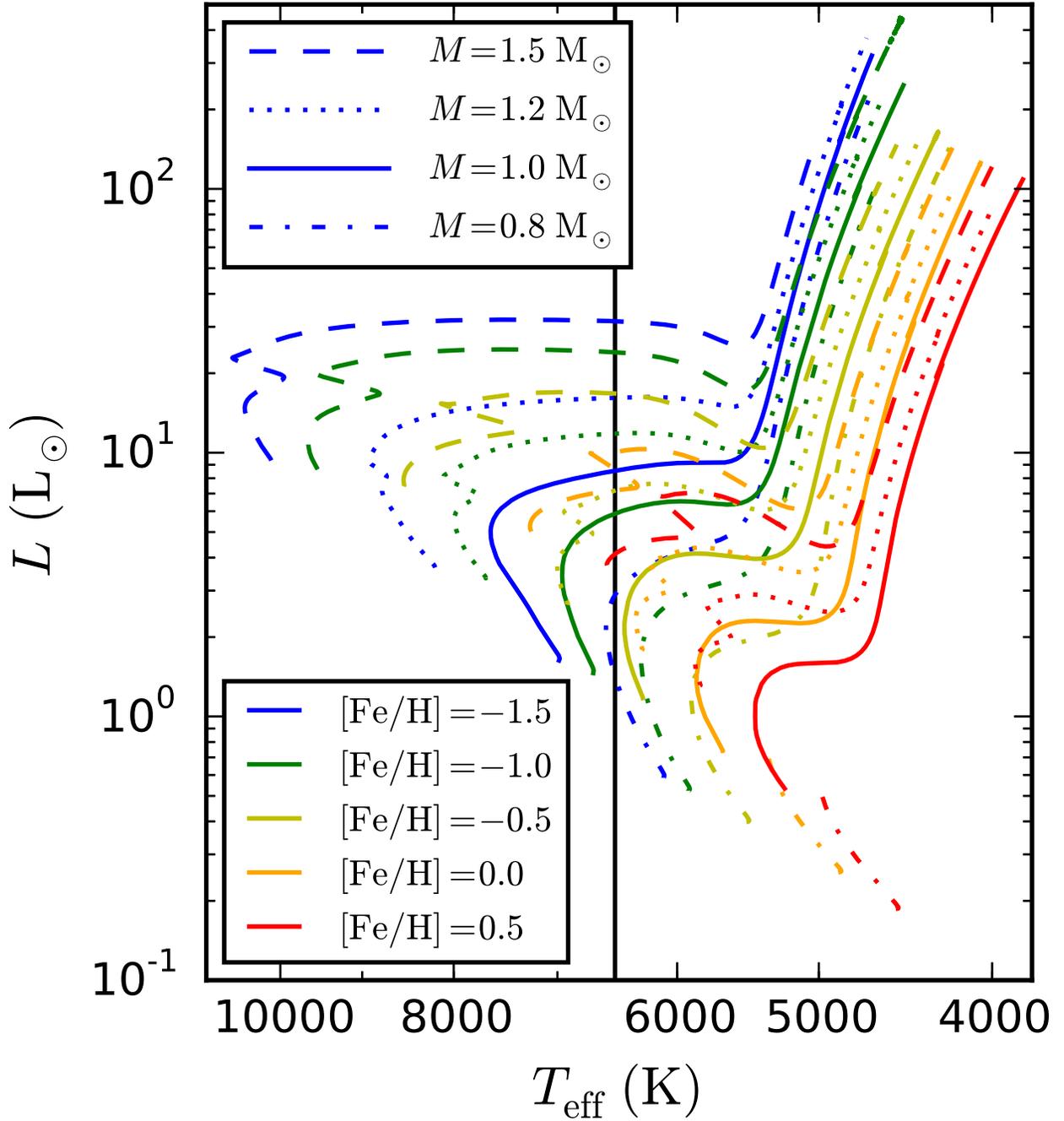}
        \caption{The evolutionary tracks from which models were picked. The different line styles refer to different masses and different colors to metallicity. We only used models with $\teff < 6500$ K (to the right of the vertical, black line).}
        \label{fig:hrdiag}
\end{figure}

The basic set of models was constructed with the Yale stellar evolution code, YREC \citep{Demarque2008}. We used the OPAL equation of state \citep{Rogers2002},  OPAL high temperature opacities \citep{Iglesias1996}, and the \citet{Ferguson2005} low temperature opacities. Nuclear reaction rates were from \citet{Adelberger1998}, except for the $^{14}$N($p,\gamma$)$^{15}$O reaction, for which we used the  \citet{Formicola2004} result. Unlike standard solar models, we did not include diffusion or core overshoot. The initial models were constructed with Eddington $T$-$\tau$ relationship and the solar calibrated value of $\alpha$, which is 1.69 in the absence of diffusion. These models were constructed for four masses ($M=0.8, 1.0, 1.2$, and $1.5$~\msun) and at five metallicities ($\feh=0.5, 0.0, -0.5 ,-1.0, and -1.5$), where we assume that $\feh=0$ corresponds to $Z/X=0.023$ \citep{gs98}. We evolved the models up the giant branch. Models were output at many different ages for the purpose of calculating pulsation frequencies. The evolutionary tracks from which our models were picked are shown in  Fig.~\ref{fig:hrdiag}. For a subset of these models, we constructed models with identical mass, radius, age, and luminosity, but with the  Krishna Swamy (henceforth, KS) $T$-$\tau$ relation using the calibration used to construct standard solar models.

A calibrated solar model is constructed by requiring a 1 \msun\ model to have the observed luminosity and radius at the solar age of 4.57 Gyr. To achieve this, one notes that there are two free parameters in the model --- the initial helium abundance, $Y_0$, and the mixing length  parameter, $\alpha$. To get a calibrated model, $Y_0$ and $\alpha$ are chosen so that the two constraints (on luminosity and radius) are satisfied at the solar age. We follow a similar procedure for models of different masses and ages.  Our ``standard'' models are the Eddington models described above, and we require the KS models to have the same radius and luminosity at the same age as the Eddington model, and this is again done by changing $Y_0$ and $\alpha$. 

The above calibration method, however, works only for a very narrow range of masses, metallicity, and ages. The main change between Eddington and KS models is the mixing length parameter $\alpha$, which is what is responsible for the surface term. However, there is a very small change in $Y_0$ too. For instance, for the Eddington solar model that we constructed, $Y_0=0.27572$, but for the KS model, it is $Y_0=0.27566$, i.e., a fractional difference of 0.02\%; both models have surface $Z/X=0.023$. Such a small change in $Y_0$ does not really change structure, except in stars with convective cores; the small difference in the size of the convective core means that the frequency differences are no longer, strictly speaking, due to a near-surface change. For low mass stars with radiative cores, the difference begins to matter in the upper part of the main sequence where the slight difference in the initial metallicity manifests as a difference in the size of the helium core, and hence deep-seated difference between the sound speed profiles of the two models. As a result, we have only a very restricted set of models for which we could successfully simulate the surface term in this manner. These successful models are listed in Table~\ref{tab:tab1}. 

The frequency differences between these KS and Eddington models essentially represent the surface term.
We assume the KS models to be the proxy stars and that they have relative frequency errors of $10^{-4}$.
We used these errors to generate 100 realizations of the frequencies of these models. Thus for
each model pair we now get 100 realizations of the surface term.
Each of these surface-term realizations
 were then fit using the different functions described in Section~\ref{sec:surf} to see 
whether they can model surface term between these models.

\subsection{Frequency response for a given near-surface differences}
\label{subsec:conv}

The theory of linear adiabatic stellar pulsations tells us that the frequency differences between two models that are not very different from each other can be expressed as:

\be
\frac{\delta\nu_i}{\nu_i}=\int K^{c^2,\rho}_i(r)\frac{\delta c^2}{c^2}(r)dr+
\int K^{\rho,c^2}_i(r)\frac{\delta\rho}{\rho}(r)dr,
\label{eq:kernels}
\ee
\noindent where $\delta c^2/c^2$ is the relative difference between the squared sound speed profiles of two models, $\delta\rho/\rho$ is the relative density difference, and the  functions $K^{c^2,\rho}_i(r)$ and $K^{\rho,c^2}_i(r)$ are known functions and can be calculated for a given model \citep[see e.g.,][]{Antia1994}. Implicit in the equation is the assumption that the models have the same mass and radius.

Equation~\ref{eq:kernels} also shows that we can calculate the frequency differences between two models by multiplying a given relative sound speed difference and/or a given relative density difference with the kernels for a model and integrating the resulting function. This is the path we take in order to simulate the surface term over a large part of the HR diagram. To this end, we calculated the kernels for all the Eddington models described in Section~\ref{subsec:calib} and applied known near-surface differences in sound speed to determine the frequency differences resulting from the difference. 

\begin{figure}
                \plotone{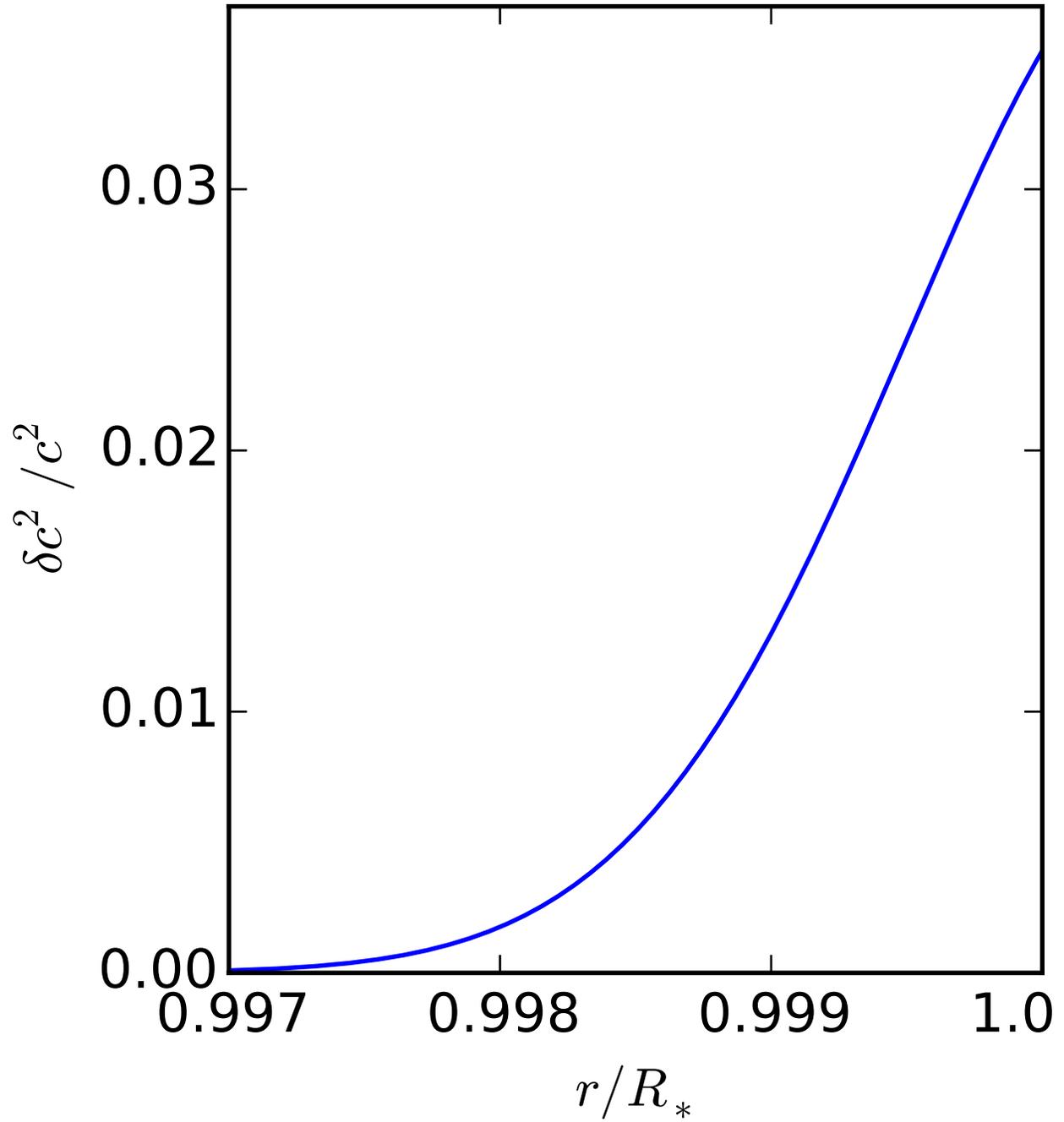}
        \caption{An artificial sound speed difference that was used to simulate a surface term using Eq.~\ref{eq:kernels}. The difference is a truncated Gaussian.}
        \label{fig:cgaus}
\end{figure}

We consider two distinct cases. In the first case, we use an artificial sound speed difference that is confined to the near-surface layers (Fig.~\ref{fig:cgaus}), assuming no density difference. The profile is basically a truncated Gaussian, and hence we call this set the ``Gaussian'' set. This is essentially a toy model since equations of
stellar structure imply that a sound-speed difference will be accompanied by a density difference. Nonetheless,
this gives us a clean signal to try out the fits.

\begin{figure}
                \plotone{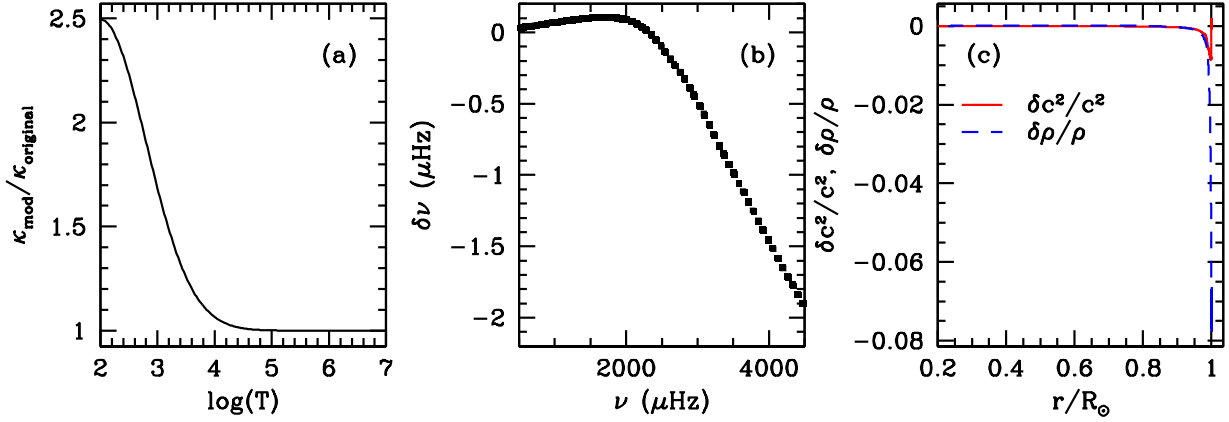}
        \caption{(a) The near-surface opacity modification applied to the solar model. (b) The frequency differences between a normal and a modified opacity solar model. (c) The relative squared sound speed difference and relative density difference between the two models. These differences were used in Eq.~\ref{eq:kernels} to simulate a surface term in different models.}
        \label{fig:art}
\end{figure}

The second case is more realistic in that we use self-consistent sound-speed
and density differences.  We constructed two solar models, one with the physics described in 
Section~\ref{subsec:calib}, and the other where the opacities were modified close to the surface
as a function of temperature (see Fig.~\ref{fig:art}a). The changes in opacity are not physically
motivated; any similar change will work for the purpose of simulating the surface term.
The frequency difference between the two solar models are shown in Fig.~\ref{fig:art}(b), 
and the relative squared sound speed and density differences are shown in Fig.~\ref{fig:art}(c). 
We then find the frequency response to these
sound speed and density differences using Eq.~\ref{eq:kernels}.
We call this the ``Opacity set''.

The frequency differences obtained for both the Gaussian and Opacity sets were then fit by the forms described in Section~\ref{sec:surf} to see which, if any, does well for which models. As before, we assumed relative frequency errors of $10^{-4}$ and generated 100 realizations of the surface term for each model by adding random errors
consistent with the assumed errors.
In order to fit the KBCD08 model, we first determined the slope $b$ from the frequency differences obtained with Eq.~\ref{eq:kernels}
by using the near-surface structural differences and the kernels of a solar model constructed with the same physics. This was done separately for the Gaussian and Opacity sets. The same frequency differences were used as the proxy solar surface term to fit the ScS-1 and ScS-2 models. Given the models range from main sequence models to red giants, we fit the surface term for different frequency ranges. For models with $\log g > 2.7$ we fit the simulated surface term in the range $\numax \pm 5\Delta\nu$. For stars with $\log g \leq 2.7$ (i.e., more evolved stars), we fit over the range $\numax \pm 2\Delta\nu$.

\section{Results}
\label{sec:res}

\subsection{Different solar models}

As had been mentioned earlier, the surface term between the Sun and solar models is not unique, but rather depends on the model. Thus the first exercise we attempted was to try to fit the different models of the surface term to the frequency differences between the Sun and different solar models.  We used five different solar models: the seismic model from \citet{antia1996} which we call INVBB, model BSB(GS98) from \citet{bahcalletal2006}, which we refer to as BSB, Model S from \citep{Christensen1996}, BP04  from \citet{bahcalletal2005}, and STD from \citet{basuetal2000}. We only used the $l=0$ modes for the fits and assumed a relative error of $1\times 10^{-4}$ on the frequencies. We fit the frequency differences in the range $\numax \pm 5\Delta\nu$.  Note that by definition the  models ScS-1 and ScS-2 will fit the data, hence we concentrate only on the other forms. 

\begin{figure*}
                \plotone{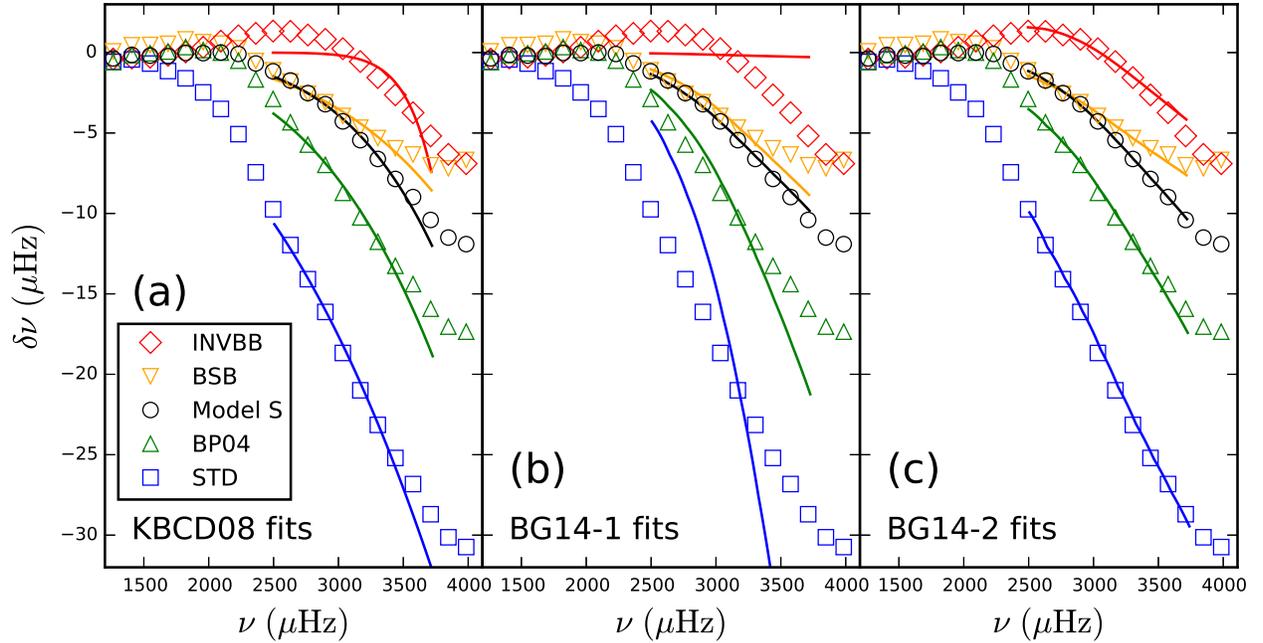}
        \caption{The surface term between the Sun and different published solar models fit with different models of the surface term in the range $\numax \pm 5\Delta\nu$. In each panel, the points are the data, and the lines are the fits. The different colors and symbols indicate different solar models.}
        \label{fig:solarfit}
\end{figure*}

Our results are shown in Fig.~\ref{fig:solarfit}.  As can be seen, the fits can be surprisingly poor even for solar models. KBCD08 was formulated with the help of Model~S, and sure enough, it fits the surface term of the model quite well, and the fit can be improved by changing the range of frequencies fit. BG14-1 is only able to properly fit the Model S surface term; it fits the surface term for INVBB extremely badly.  The most promising is BG14-2, which fits the surface term for all the models reasonably well, and the fit can be improved by changing the frequency range of the fits. This gives us an early indication of the results we can expect for simulated surface term for the non-solar models.

\subsection{Calibrated models}

\begin{figure}
                \plotone{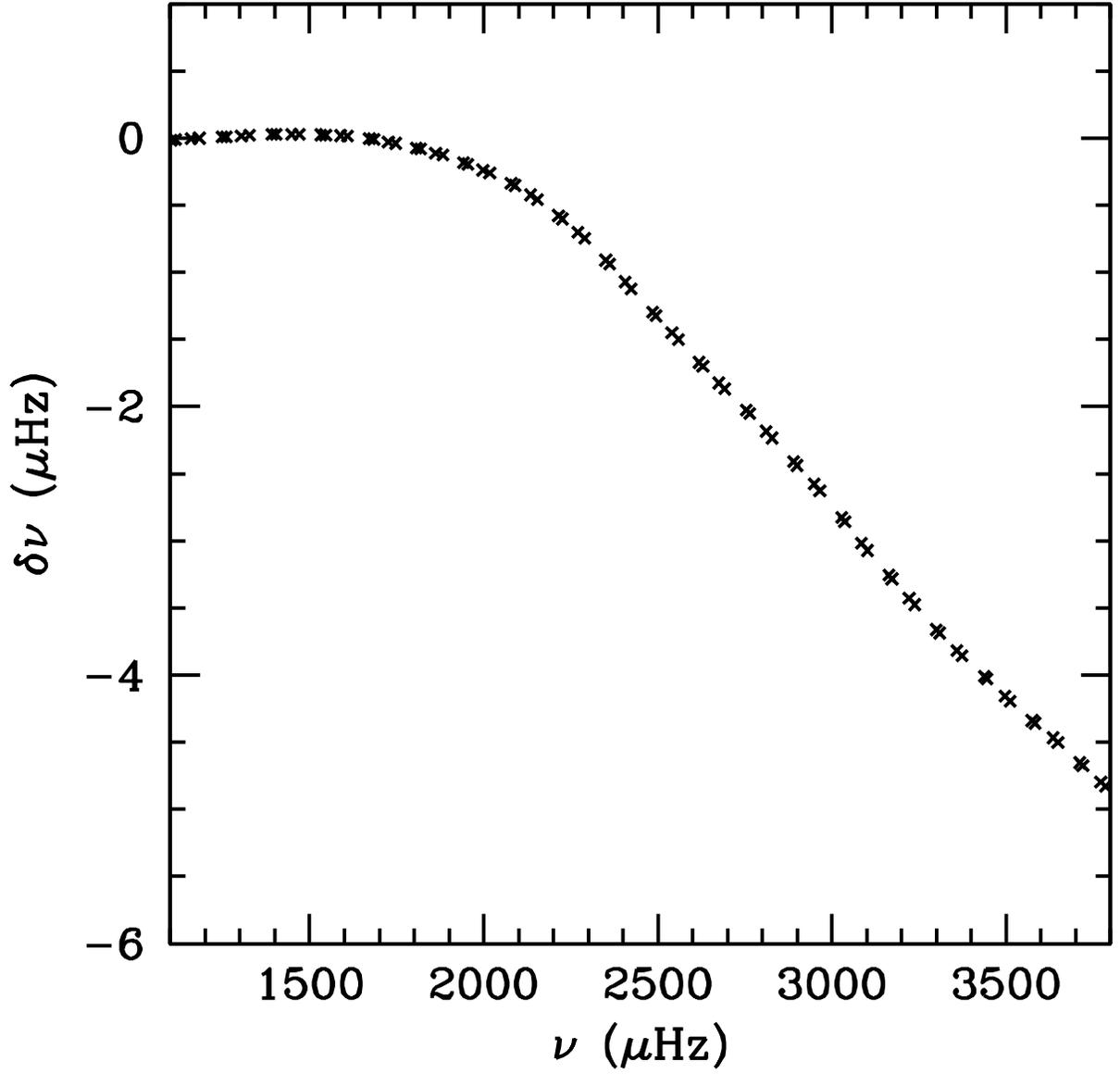}
        \caption{Frequency differences for $l \le 3$ models for a solar model constructed with the  Eddington $T$-$\tau$ relation and another constructed with the Krishna Swamy  $T$-$\tau$ relation \citet{KrishnaSwamy1966}.}
        \label{fig:eddiks}
\end{figure}

\begin{figure}
                \plotone{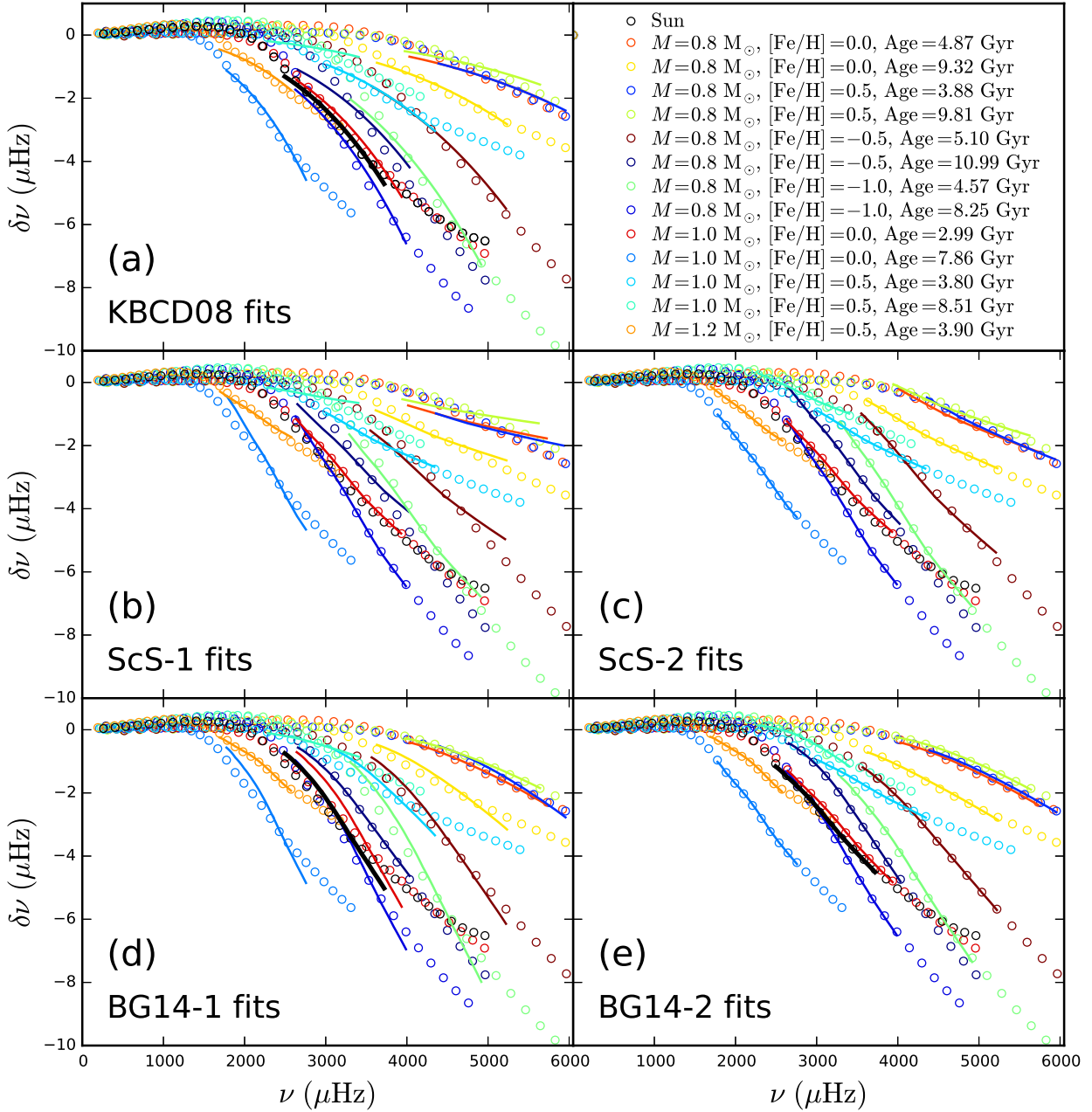}
        \caption{The frequency differences between the calibrated pairs of models fit with the five surface term models in the range $\numax \pm 5\Delta\nu$. The points are the frequency differences, and the lines are the best fit.}
        \label{fig:calib}
\end{figure}

We turn our attention next to the frequency differences between the pairs of models with identical properties except for the model atmosphere used to construct them. The fits to the frequency differences between the thirteen pairs of models is shown in Fig.~\ref{fig:calib}. The fit is only shown for error-free realization of the
surface term in each case.
For model KBCD08, the power $b$ in Eq.~\ref{eq:KBCD08} was determined from the frequency differences between two solar models that were constructed with different model atmospheres (Fig.~\ref{fig:eddiks}) and then applied to the other differences. Since errors in solar oscillation frequencies are very small,
a few parts in $10^6$ \citep[see, e.g., Table~1 of][]{chaplinetal2007}, we determined
$b$ from error-free data.  For models ScS-1 and ScS-2, the frequency differences shown in Fig.~\ref{fig:eddiks} were assumed to represent the solar surface term, and that, in the scaled form, was applied to the other differences. Again, the error-free version was used for this.

\begin{figure}
                \plotone{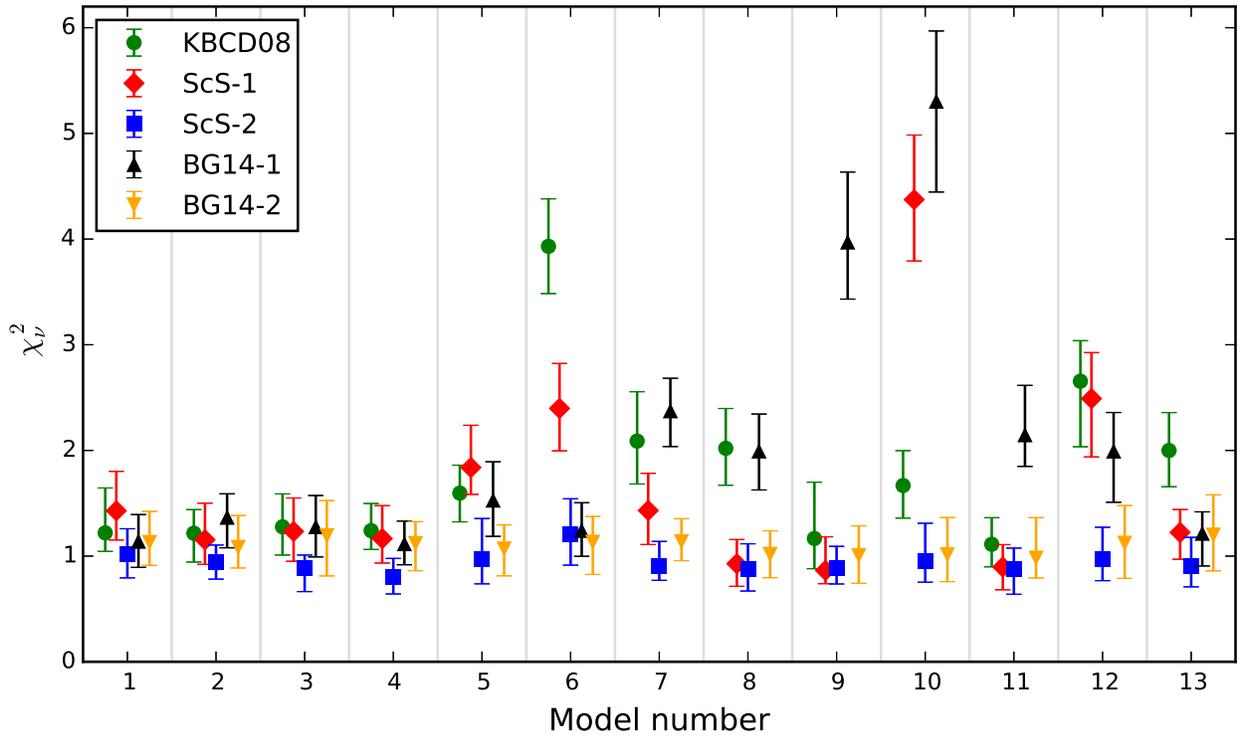}
	\caption{The median $\chired$ with $1\sigma$ spread from 100 realizations of simulated measurement errors for the fits shown in Fig.~\ref{fig:calib}. The vertical, gray lines separate the different models.  The model numbers correspond to the those listed in Table~\ref{tab:tab1}.}
      \label{fig:chi_calib}
\end{figure}

In Fig.~\ref{fig:chi_calib}, we show the goodness of fit in terms of a reduced $\chi^2$ ($\chired$) for the five methods.  
The quantity $\chired$ is defined in the conventional manner to quantify the mismatch between
the data and the model, i.e.,
\begin{equation}
\chi_\nu=\frac{1}{N-m-1}\sum_{i=1}^N \frac{ (\delta\nu_{i,{\rm surf}}-\delta\nu_{i,{\rm fit}})^2}{\sigma_i^2},
\label{eq:chisq}
\end{equation}
where, $\delta\nu_{i,{\rm surf}}$ is the surface term we are trying to fit, $\delta\nu_{i,{\rm fit}}$ is
the surface term model (KBCD08, ScS-1, ScS-2, BG14-1 or BG14-2) at frequency $\nu_i$, $\sigma_i$ 
is the  uncertainty in $\nu_i$, $N$ is the total number of points fitted and
$m$ the number of free parameters. As usual, the smaller the value of $\chired$, the better the fit to
the surface term.
What we have plotted is the median value of $\chired$ obtained from the 100 realizations for each
model pair. The vertical ``error bars'' are the $1\sigma$ spread in $\chired$ obtained for the
100 realizations.
The results are clear.  The one-term fitting models, KBCD08, ScS-1, and BG14-1, can perform fairly 
poorly for many of the stellar models.  The two two-term fits, on the other hand, both perform spectacularly well for these thirteen calibrated models.  The extra term in the ScS-2 and the 
BG14-2 fitting methods appear to be needed to improve the fits.

\subsection{Surface term from known near-surface differences}

\begin{figure}
                \plotone{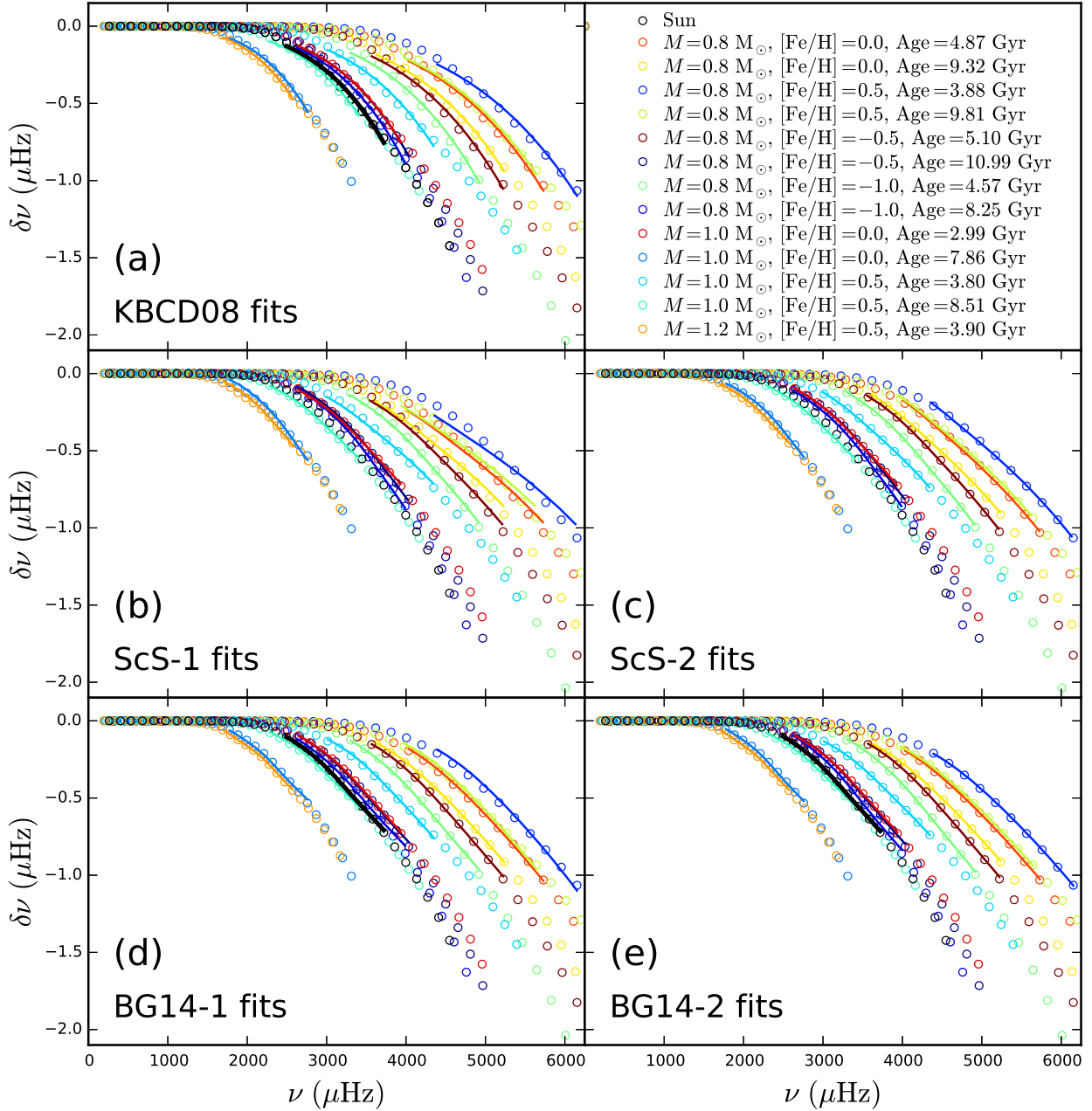}
        \caption{The same as Fig.~\ref{fig:calib}, but for the case in which the surface term was simulated by determining
the frequency response of a sound-speed perturbation in the form of a truncated Gaussian (Fig.~\ref{fig:cgaus}).}
        \label{fig:gauall}
\end{figure}

In Fig.~\ref{fig:gauall}, we show the fits to the surface term for the Gaussian set for the same subset of models shown in Fig.~\ref{fig:calib} and listed in Table~\ref{tab:tab1}. 
The one-term fits perform  better for the Gaussian set than the calibrated set, though again
the two-term fits do better, but only marginally so in this case. 

\begin{figure}
\epsscale{0.9}
                \plotone{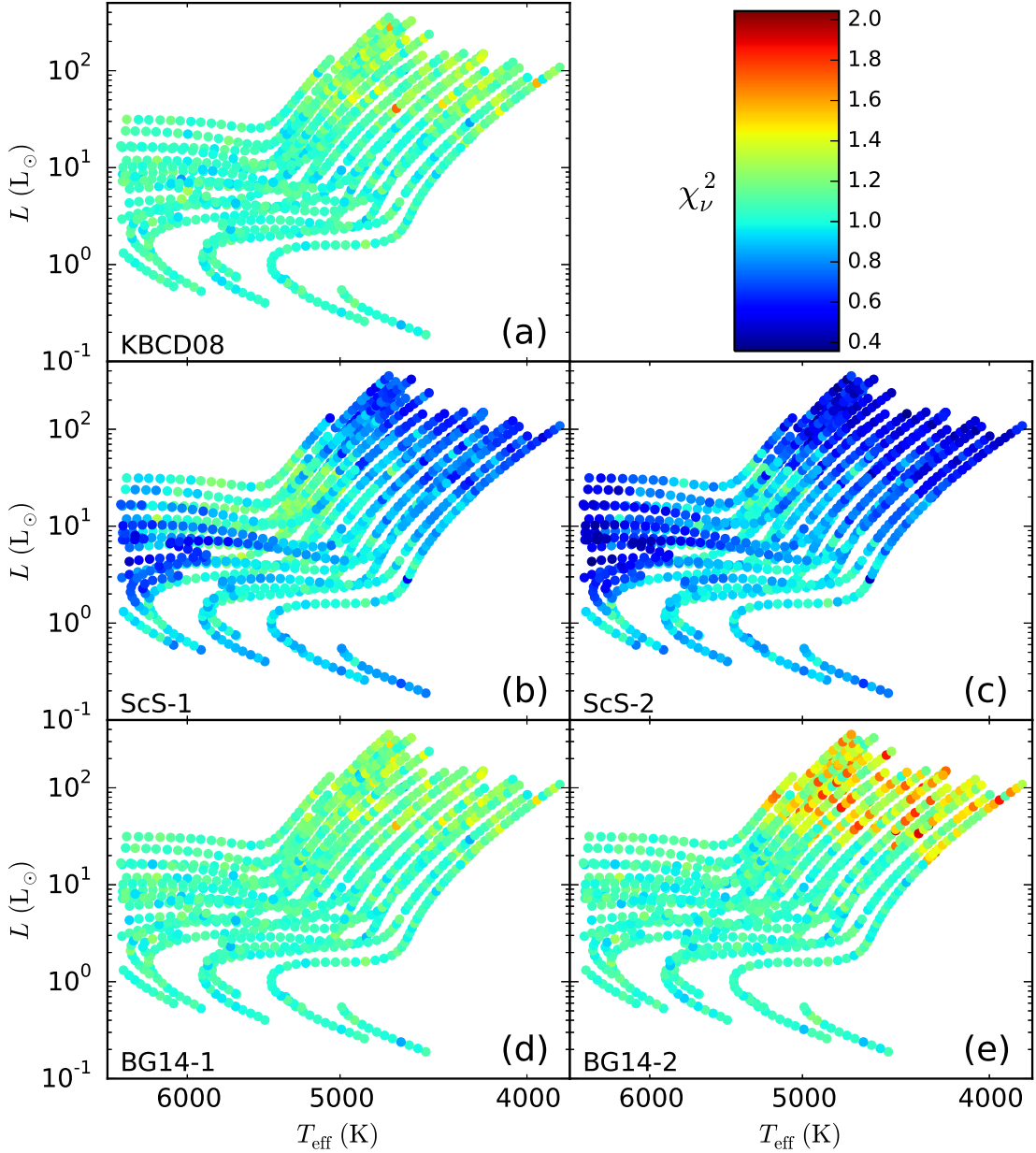}
        \caption{The median $\chired$ of each stellar model from 100 realizations of measurement errors for the different surface term models over the HR diagram for the case in which the surface term was simulated by determining the frequency difference caused
by an artificial sound speed difference in the form of a truncated Gaussian (Fig.~\ref{fig:cgaus}).  The discontinuity in color in the giant branch, most easily seen in the BG14-2 panel, is the switch from changing the fitting range from $\numax \pm 5\Delta\nu$ to $\numax \pm 2\Delta\nu$.  BG14-2's red branch fits also appear to be moderately worse than BG14-1's. However, this is because the extra fitting parameter does not significantly improve the fit, and the $\chired$ penalizes fitting formulas with more free parameters.}
        \label{fig:gauss}
\end{figure}

As mentioned earlier, we also simulated the surface term for a large number of models, and the models
shown in Fig.~\ref{fig:gauall} is only a small subset of all the models.
The results for the entire set of 1321 models is shown in Fig.~\ref{fig:gauss}. We have color-coded the points 
with the median value of $\chired$ obtained for the 100 realizations of the surface term at given points
on the HR diagram. We see that most models work reasonably well in this case, the ScS-2 doing marginally
better, particularly in the red-giant range.

\begin{figure}
                \plotone{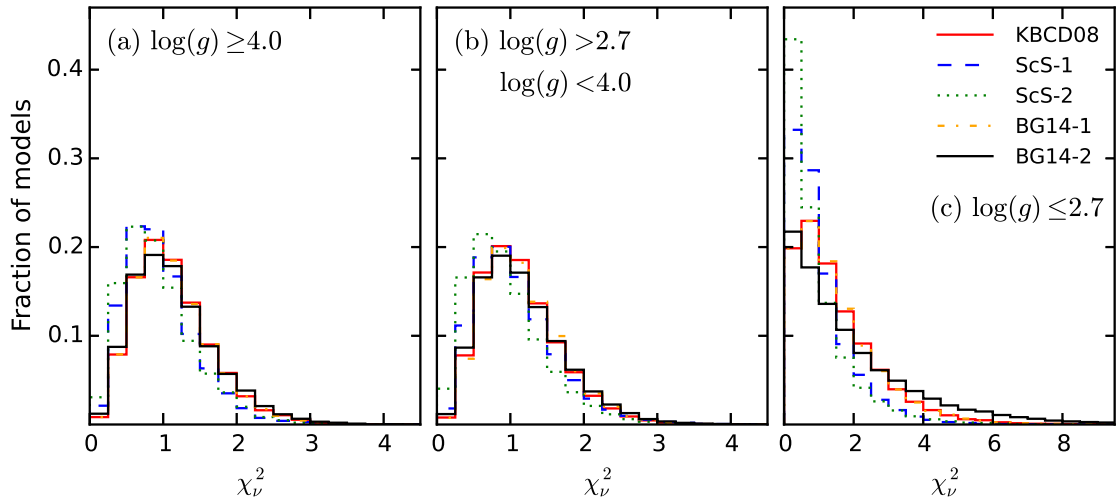}
	\caption{The distribution of $\chired$, using all 100 realizations of each stellar model, for the different surface term models for three ranges of $\log g$ for the Gaussian set. Note the change in the scale of the x-axis for the low $\log g$ panel.}
      \label{fig:chi_gauss}
\end{figure}

We show the the results more quantitatively in Fig.~\ref{fig:chi_gauss}, where we have plotted the distribution of $\chired$ for all realizations and all stellar models for each of the five surface-term models. 
Since it is clear the results depend on the evolutionary state of the star, 
we have divided our sample into three bins in $\log g$. 
The ScS-2 model appears to work the best in all cases, but the others are not too bad either.

\begin{figure}
                \plotone{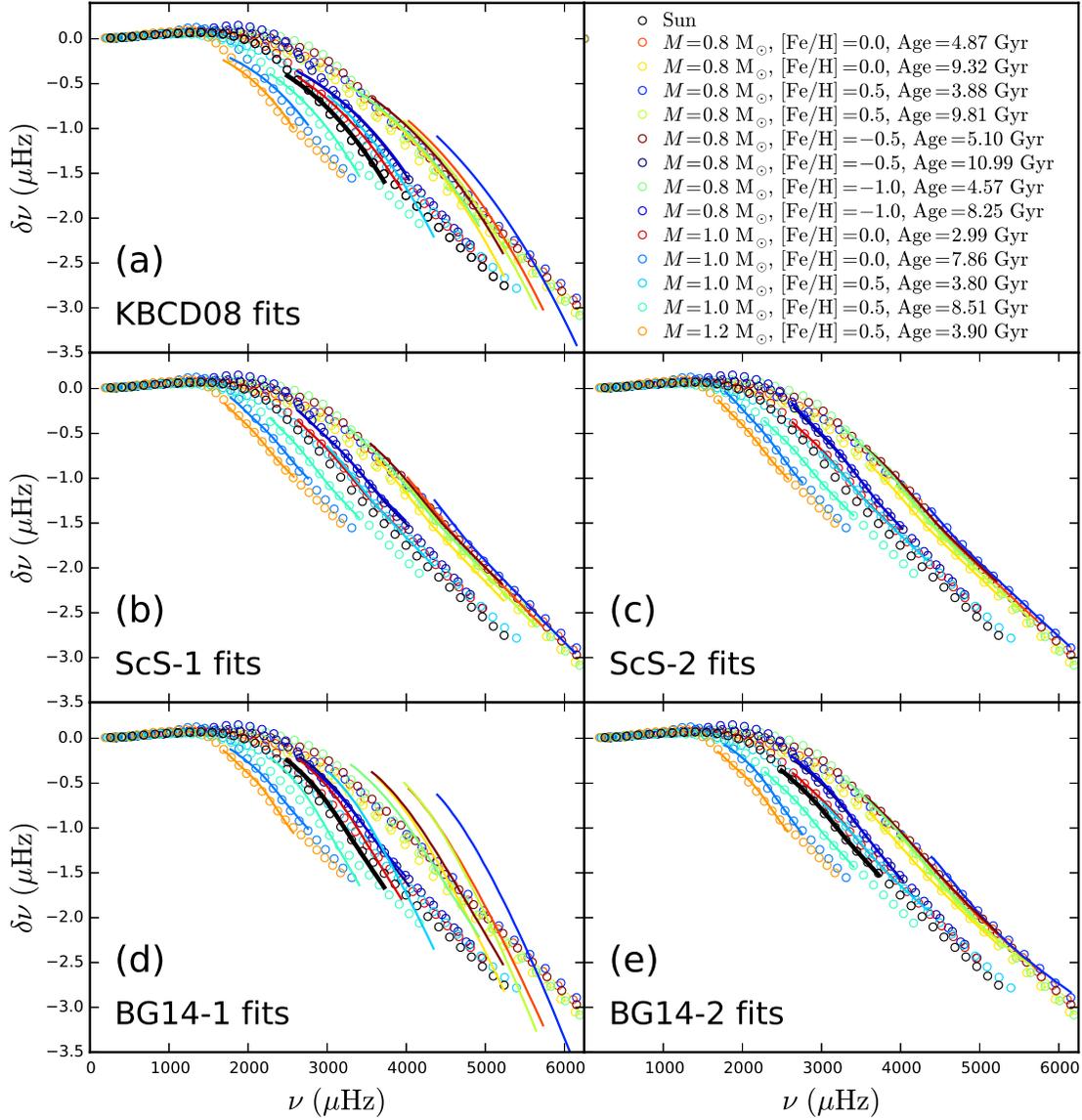}
        \caption{The same as Fig.~\ref{fig:calib}, but for the case in which the surface term was simulated using sound speed and density differences derived from near-surface changes to opacity  shown in Fig.~\ref{fig:art}.}
        \label{fig:opacall}
\end{figure}

\begin{figure}
                \plotone{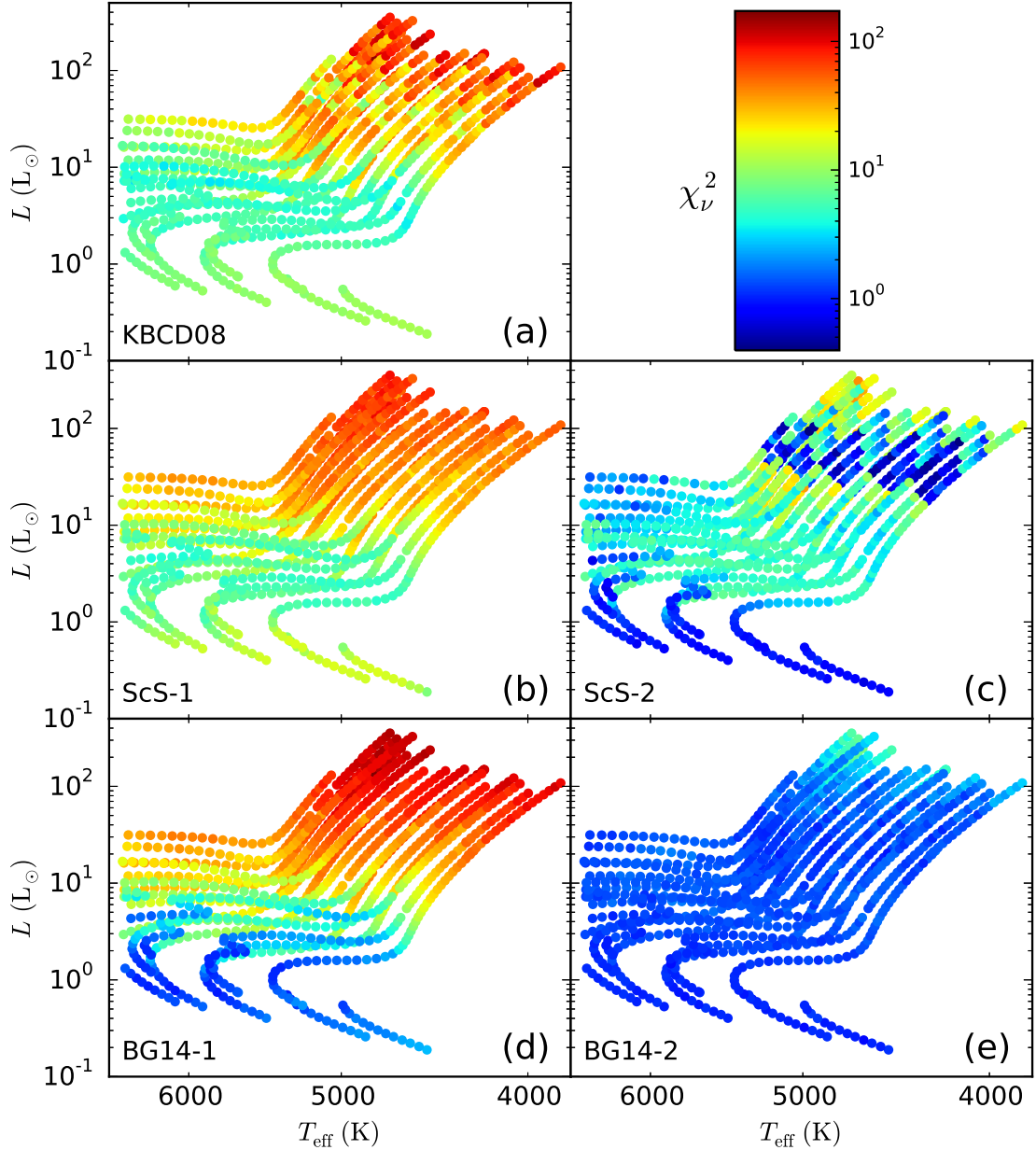}
        \caption{The same as Fig.~\ref{fig:gauss}, but for the case in which the surface term was simulated by using sound speed and density differences derived from near-surface changes to opacity  shown in Fig.~\ref{fig:art}.  Note the the color scale extends to a much higher value of $\chired$ than Fig.~\ref{fig:gauss} and is now logarithmic.}

        \label{fig:opac}
\end{figure}

Given that the surface term simulated by determining the frequency response to a sound speed perturbation in the form of a truncated Gaussian is not completely realistic, since we assume that there is a sound speed difference in the absence of a density difference, we turn our attention to the results for the Opacity set. In Fig.~\ref{fig:opacall}, we show the fits to 
the subset of models listed in Table~\ref{tab:tab1}, and Fig.~\ref{fig:opac} shows how the fits behave over the HR diagram. It is
clear that, in this more realistic surface term, BG14-2 works better than the rest. ScS-2 also works  well.
Neither ScS-1 nor KBCD08 are able to fit the frequency differences. BG14-1 works in the main sequence, 
but not for evolved stars.

\begin{figure}
                \plotone{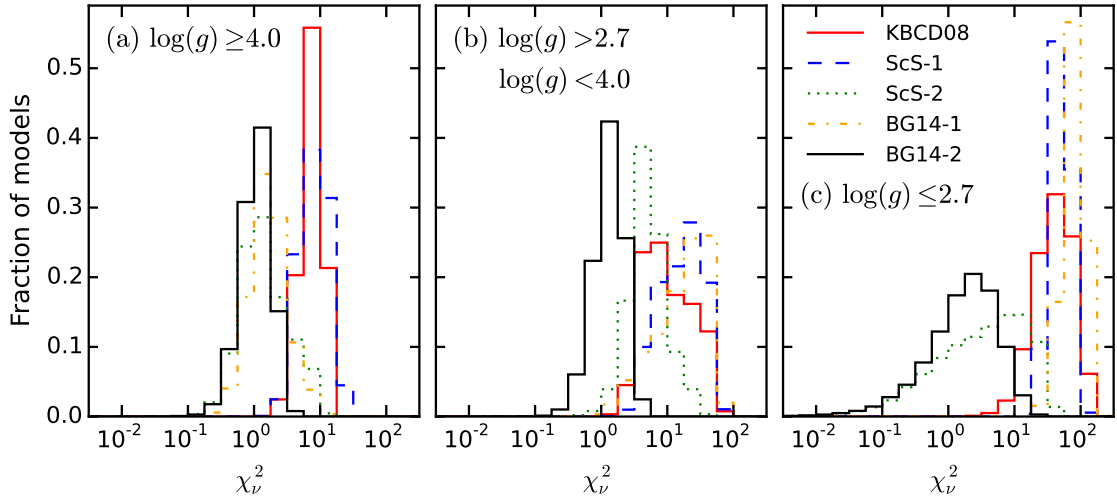}
	\caption{The distribution of $\chired$, using all 100 realizations of each stellar model, for the different surface term models for three ranges of $\log g$ for the Opacity set. Note that the x- and y-scales have changed from Fig.~\ref{fig:chi_gauss} and particularly that the x-scale is now logarithmic.}
      \label{fig:chi_opac}
\end{figure}

In Fig.~\ref{fig:chi_opac} we show the $\chired$ distribution for the entire sample 
in the same manner as for the Gaussian set. The two two-term models perform the best 
for main sequence stars. The three one-term models struggle. While BG14-1
works in the main sequence, it quickly deteriorates. KBD08 and ScS-1 so badly everywhere.
Both two-term 
models do well in the main sequence, BG14-2 being slightly better. In the sub-giants 
segment, we start to see BG14-2 pull away from the crowd, while the giants undeniably show 
that BG14-2 is by and far the best surface term model for low $\log g$ stars.

\section{Discussion and Conclusion}

The surface term --- the frequency differences between a star and its model caused by our inability to properly model the near-surface layers --- makes asteroseismic analyses difficult. The near-surface contribution is usually removed using {\it ad hoc} models of what the difference should look like. In this work, we have simulated different forms of the surface term for models that span a large part of the HR diagram and tested different models for removing the surface term. 

There are issues with many of the surface-term models even for main sequence stars, and
most models perform very badly in the sub-giant and red giant region. 
However, the recently proposed two-term model of \citet{Ball2014}\ works well in all the parts of the HR diagram.
The scaled solar model with an offset (ScS-2) works reasonably on the main sequence
but not elsewhere.
We, therefore, conclude that the two-term model of \citet{Ball2014}\ (Eq.~\ref{eq:bg142} of this paper) is the model that should be used for future asteroseismic analyses.

Our results are not affected by the fact that we have chosen a uniform relative frequency errors.
We have repeated the test on a subset of our models using an error-distribution that has
minimum errors at $\nu_{\rm max}$ and increasing errors away from $\nu_{\rm max}$. Those
results are completely consistent with our previous results.

\begin{figure*}
\epsscale{0.8}
                \plotone{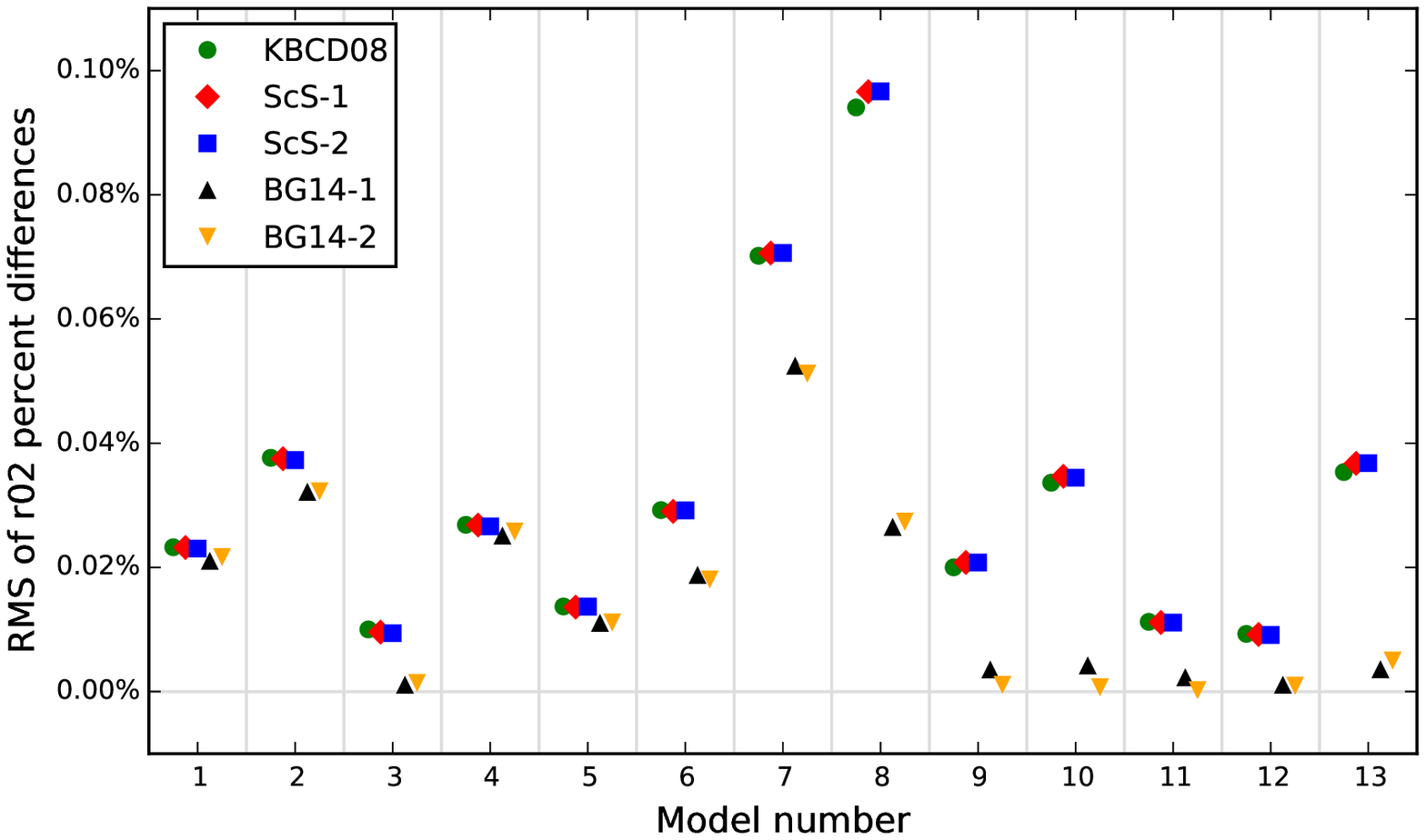}
	\caption{The RMS difference, expressed in percentages, between the separation ratio $r_{02}$
calculated using surface-term removed frequencies and $r_{02}$ calculated using the original frequencies
of the models listed in Table~\ref{tab:tab1} constructed with KS atmospheres. The RMS 
differences are plotted against model number. The vertical gray lines separate the models.
We plot the differences for all the different surface-term models that we have tested in this work. For reference, typical uncertainties in calculating the $r_{02}$ (and $r_{01}$ and $r_{10}$) separation ratios are of order a few percent.}
      \label{fig:r02}
%\end{figure*}
%
%\newpage
%
%\begin{figure*}
                \plotone{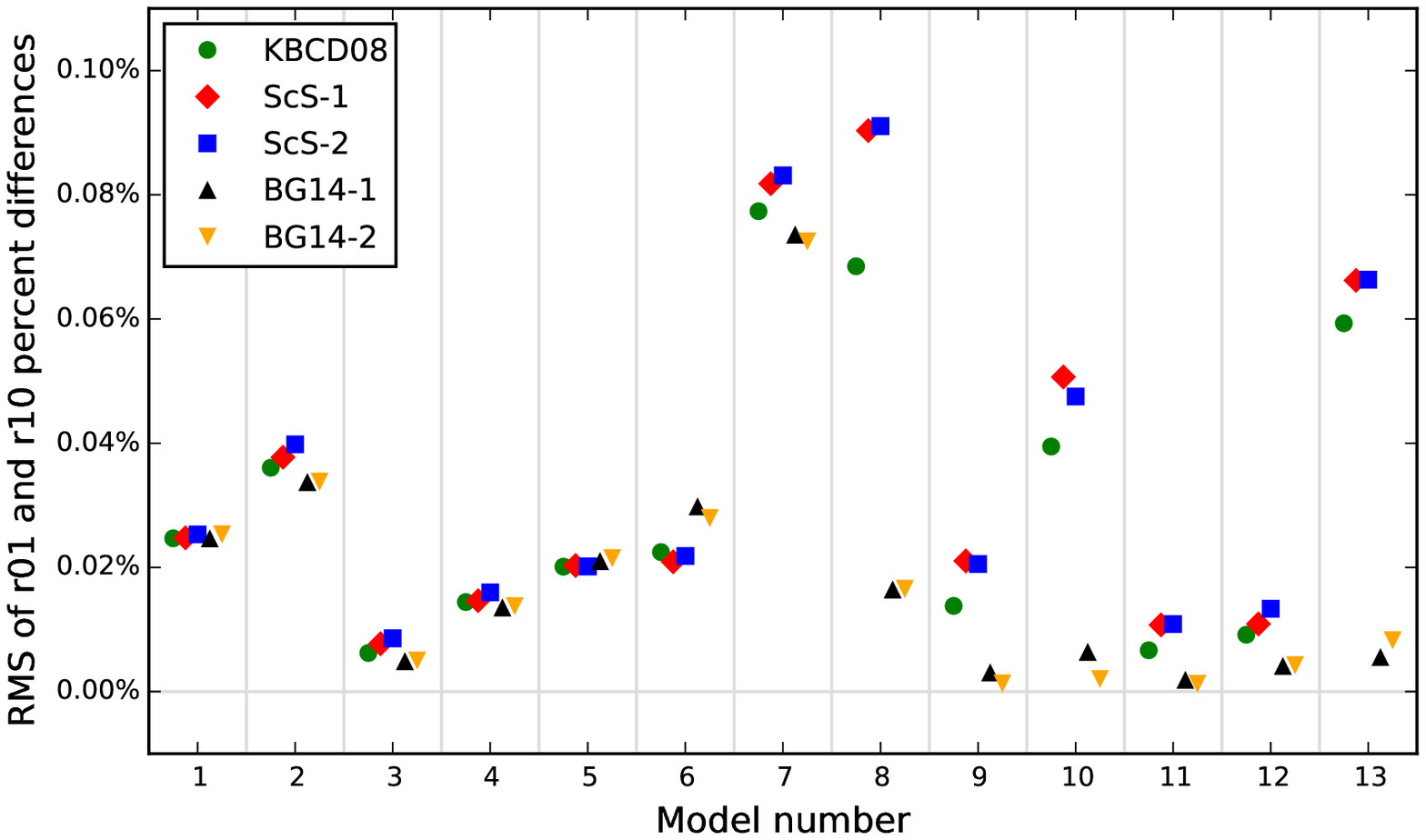}
	\caption{The same as Fig.~\ref{fig:r02}, but for the frequency ratios $r_{01}$ and $r_{10}$.
We have calculated these ratios as per the definitions in \citet{Silva2011}.}
      \label{fig:r01}
\end{figure*}

It should be noted surface term corrections can in many cases distort or hide the seismic
signature from the stellar interior. As a result, stellar models with different properties can
fit the frequencies of the same star after surface term corrections.  The way to way to distinguish these models is to use frequency separation ratios.  Frequency separation ratios
are essentially insensitive to the surface term, and hence can be calculated with the
uncorrected frequencies of the model. There have not been any tests of how models BG14-1 or BG14-2
could distort the signature from stellar interiors. To test this we makes use of the
model pairs listed in Table~\ref{tab:tab1}. Recall that we had constructed pairs of models
with different atmospheres; the frequency differences between the two models is the surface term.
We fit this surface term and subtracted it out from the models with the KS atmospheres.  We then
used these `corrected' frequencies to calculate different frequency ratios and
compared them with the ratios calculated with the original frequencies. In Fig.~\ref{fig:r02} and
Fig.~\ref{fig:r01} we show root-mean-square (RMS) differences (in percent) between the ratios
calculated with the surface-term removed frequencies and those calculated with the
original frequencies. We find that the BG14-1 and BG14-2 surface-term  models do as well, and 
sometimes even better, than the other surface-term models that we have used.  Thus, the
adoption of this correction poses no more risk than adopting any of the others.

Most of the results, except of course the ones in Fig.~\ref{fig:r02}\ and Fig.~\ref{fig:r01},
were obtained using radial modes only. The conventional practice has been to define the surface
term using radial modes and then apply the correction to the non-radial modes after 
scaling with $Q_{nl}$. The explicit presence of the factor $E_{nl}$ in BG14-2 should have,
in principle, allowed us to define the surface term using both radial and non-radial models.
However, for the few cases that we tested, we find that even for the BG14-2 model, the best results are still obtained
by first defining the surface term using radial modes and then applying them to the
non-radial ones. This is the case for stars beyond the main sequence. For main-sequence
stars, it did not matter whether we used just the radial modes or all modes.

\acknowledgments We than the anonymous referee whose comments and suggestions have led to improvements in this paper.
This work was supported by NSF grant AST-1105930 and NASA grant NNX13AE70G to SB. JRS was supported from NASA ADAP grant NNH14ZDA001N.

%\bibliography{ms}

\newpage

%%%%%%%%%%%%%%%%%%%%%%%%%%%%%%%%%%%%%%%%%
%			TABLES
%%%%%%%%%%%%%%%%%%%%%%%%%%%%%%%%%%%%%%%%%

\begin{deluxetable}{cccccc}
\tablecolumns{5}
\tablecaption{Properties of models that could be successfully constructed with two different atmospheric models}
\tablehead{
\colhead{Model number}&\colhead{Mass (\msun)}& \colhead{Radius (\rsun)}& \colhead{$T_{\rm eff}$ (K)}& \colhead{Age (Gyr)}&
\colhead{[Fe/H]}
}
\startdata
1  & 0.8 & 0.746 &  4944 &   4.87 &   0.0  \\
2  & 0.8 & 0.775 &  5043 &   9.32 &   0.0  \\
3  & 0.8 & 0.727 &  4557 &   3.88 &   0.5  \\
4  & 0.8 & 0.754 &  4650 &   9.81 &   0.5  \\
5  & 0.8 & 0.757 &  5616 &   5.10 &   -0.5 \\
6  & 0.8 & 0.857 &  5800 &  10.99 &   -0.5 \\
7  & 0.8 & 0.765 &  6054 &   4.57 &   -1.0 \\
8  & 0.8 & 0.857 &  6183 &   8.25 &   -1.0 \\
9  & 1.0 & 0.963 &  5754 &   2.99 &   0.0  \\
10 & 1.0 & 1.162 &  5852 &   7.86 &   0.0  \\
11 & 1.0 & 0.946 &  5325 &   3.80 &   0.5  \\
12 & 1.0 & 1.063 &  5422 &   8.51 &   0.5  \\
13 & 1.2 & 1.325 &  5812 &   3.90 &   0.5  \\
\enddata
\label{tab:tab1}
\end{deluxetable}

%%%%%%%%%%%%%%%%%%%%%%%%%%%%%%%%%%%%%%%%%
%			Figures
%%%%%%%%%%%%%%%%%%%%%%%%%%%%%%%%%%%%%%%%%

\newpage

\newpage

\newpage

\newpage

\newpage

\newpage

\newpage

\newpage

\newpage

\newpage

\newpage

\newpage

\newpage

\newpage

\newpage

\end{document}